# Estimating Deprivation Cost Functions for Power Outages During Disasters: A Discrete Choice Modeling Approach

Xiangpeng Li[1], Mona Ahmadiani[2], Richard Woodward[2], Bo Li[1], Arnold Vedlitz[3], Ali Mostafavi[1]

[1] Ph.D. student, UrbanResilience.AI Lab, Zachry Department of Civil and Environmental Engineering, Texas A&M University, College Station, TX, 77843, USA; e-mail: xplli@tamu.edu

[2] Assistant Professor, Department of Agricultural Economics, Texas A&M University, College Station, TX, 77843, USA; e-mail: mona.ahmadiani@ag.tamu.edu

[2] Professor, Department of Agricultural Economics, Texas A&M University, College Station, TX, 77843, USA; e-mail: r-woodward@tamu.edu

[1] Ph.D. student, UrbanResilience.AI Lab, Zachry Department of Civil and Environmental Engineering, Texas A&M University, College Station, TX, 77843, USA; e-mail: libo@tamu.edu

[3] Professor, Institute for Science, Technology, and Public Policy, Bush School of Government and Public Service, Texas A&M University, College Station, TX, 77843, USA; e-mail: avedlitz@tamu.edu

[1] Professor, Urban Resilience.AI Lab Zachry Department of Civil and Environmental Engineering, Texas A&M University, College Station, TX, 77843, USA; e-mail: amostafavi@civil.tamu.edu

## Abstract

Systems for the generation and distribution of electrical power represents critical infrastructure and, when extreme weather events disrupt such systems, this imposes substantial costs on consumers. These costs can be conceptualized as deprivation costs—an increasing function of time without service—quantifiable through individuals' willingness to pay for power restoration. Despite widespread recognition of outage impacts, a gap in the research literature exists regarding the systematic measurement of deprivation costs. This study addresses this deficiency by developing and implementing a methodology to estimate deprivation cost functions for electricity outages, using stated preference survey data collected from Harris County, Texas. This study compares multiple discrete choice model architectures, including multinomial logit and mixed logit specifications, as well as models incorporating Box–Cox and exponential utility transformations for the deprivation time attribute. The analysis examines heterogeneity in deprivation valuation through sociodemographic interactions, particularly across income groups. Results confirm that power outage deprivation cost functions are convex and strictly increasing with time. Additionally, the study reveals both systematic and random taste variation in how individuals value power loss, highlighting the need for flexible modeling approaches. By providing both methodological and empirical foundations for incorporating deprivation costs into infrastructure risk assessments and humanitarian logistics, this research enables policymakers to better quantify service disruption costs and develop more equitable resilience strategies.

## 1. Introduction

Major disasters—such as wildfires, tornadoes, hurricanes, tropical storms, and floods—frequently damage transmission lines, substations, and fuel distribution networks, disrupting or severely constraining the supply of electricity[1,2]. For example, Hurricane Beryl on July 8, 2024 caused widespread power disruptions across the Houston metropolitan area, leaving more than 2 million customers without power for several days, resulting in more than 143 million total customer-outage hours[3]. These interruptions affect the bedrock societal functions: economic productivity; public safety; and routine functions of households, businesses and public institutions[4,5]. This breakdown in energy delivery systems restricts residents' access to essential services, including communication and healthcare[6]. Given these conditions, individuals face a critical decision: whether to risk an uncertain period of deprivation or to invest in costly backup systems to restore power independently if an outage occurs[7]. If the primary grid is either partially or completely disabled in the aftermath of such events, the demand for alternative and backup power solutions increases sharply[8,9]. In environments characterized by indeterminate power restoration timelines and limited auxiliary power, determining equitable and effective electricity distribution presents a multifaceted ethical and logistical challenge[10]. This challenge involves establishing priorities between critical infrastructure, such as hospitals and emergency shelters, and vulnerable households, including those with elderly individuals or young children[11]. Furthermore, the impacts households experience and how they respond to service interruptions vary across groups based on income, or other socio-demographic charateristics[12]. Quantification of power deprivation costs will provide valuable information to inform investments to improve the resilience of critical infrastructure improvements in hazard-prone areas, and other risk reduction efforts. This research directly addresses this gap by developing methodology to quantify these social costs of infrastructure disruption over time.

Despite a growing recognition that disasters disproportionately burden society's most vulnerable groups, most mainstream infrastructure-risk analysis appraises power-system investments almost exclusively through direct repair costs and avoided physical damage[13–15]. This misses the social fallout of prolonged outages—lost income[16], compromised health, parental stress[17], food spoilage[18], educational disruption—remains largely invisible in benefit–cost calculations because the field lacks an accepted *monetary* metric for these harms. Existing "capability," "hardship," or "well-being" lenses have deepened our qualitative understanding of human suffering during service interruptions, yet they yield ordinal or subjective scores rather than the dollar-valued user costs that standard economic appraisal frameworks require[14,19]. As a result, agencies and utilities cannot place social impacts on the same ledger as physical damage when ranking grid-hardening projects, setting restoration priorities, or defending rate-based investments. The absence of defensible, dollar-denominated *deprivation cost* functions has, therefore, become a critical blind spot: without them, resilience projects that would sharply reduce human welfare burdens often appear "uneconomic" under conventional cost-benefit thresholds, leading to deferred upgrades, slower restorations, and avoidable social harm[20,21]. By measuring the time households spend without electricity in terms of willingness-to-pay, our study closes this methodological gap and supplies the missing social-cost term that can be seamlessly embedded in engineering economics, optimization models, and regulatory impact statements.

Storm-driven outages expose an often-ignored layer of disaster loss: the deprivation cost borne by households that must function without essential services while the grid is dark[22]. Physical damage—broken lines, spoiled inventory, lost wages—are routinely monetized and thus dominate the post-event resource-allocation calculus[23]. Yet poorer neighborhoods, already lacking protective infrastructure, endure a second, less visible toll: missed dialysis sessions, discarded medications[24], sleepless nights in heat, the cognitive load of scrambling for fuel or child-care[25], and the anxiety that accompanies each uncertain hour of darkness[26]. Traditional proxy-based measures have long dominated humanitarian logistics research[27–29], though they fail to adequately capture human suffering during service disruptions[30]. Existing social-impact lenses (capability, hardship, well-being, and allied frameworks) have sharpened our qualitative understanding of this toll, but they stop short of producing the dollar-valued user costs that engineering-economic appraisals require. Without such metrics, grid-hardening or fast-restoration projects that would avert profound human suffering appear noncompetitive under conventional benefit-cost tests, perpetuating under-investment and widening inequities. To address this limitation, Holguín-Veras pioneered the concept of deprivation cost for water distribution, establishing a quantifiable function that could be integrated into humanitarian logistics mathematical models[31]. Following this foundational work, researchers have expanded deprivation cost applications across multiple domains. Numerous studies have explored deprivation costs for essential resources including food and water[27,32–35], while others have specifically examined medical service deprivation during emergencies[33,36–39]. However, no study has translated prolonged infrastructure outages themselves into empirically grounded cost functions. Infrastructure service disruptions, particularly power system failures during disasters, generate substantial yet poorly quantified social costs. While existing research has applied the travel cost method (TCM) to evaluate service accessibility and capture the burdens vulnerable communities bear during disruptions[16], and willingness-to-pay (WTP) approaches have assessed users' valuations of service continuity[19], these methods remain incomplete in estimating the welfare with respect to time deprived. The inability to quantify social user costs from infrastructure disruptions creates a significant gap in resilience investment analysis. Without measuring these human welfare impacts, decision-makers lack complete benefit-cost assessments for infrastructure improvements, systematically undervaluing resilience interventions. This analytical deficiency delays critical investments in hazard-prone areas, perpetuating infrastructure vulnerability and compounding future disaster impacts, particularly for disadvantaged communities.

To address the critical gap in the current literature on deprivation cost functions (DCFs) for infrastructure systems, this study extends beyond traditional humanitarian logistics applications to examine the user costs of the infrastructure disruptions in the U.S. Infrastructure failures generate deprivation—the acute absence of essential services—which consequently produces human suffering and economic consequences far exceeding direct monetary losses. The DCF framework fundamentally transforms infrastructure resilience investment evaluation by quantifying welfare losses through economic valuation, thereby capturing the full spectrum of societal impacts. Since resilience investments primarily benefit society through preventing substantial user costs, their accurate quantification consequently becomes essential for developing effective risk reduction strategies[40]. Employing stated-preference survey data from Harris County, Texas about willingness-to-pay for immediate power restoration, this study develops a robust methodology for estimating DCFs specifically for electrical power deprivation by utilizing discrete choice models. This approach models individuals' preferences between purchasing immediate backup power at user costs versus awaiting restoration, thereby allowing for both linear and non-linear DCFs based on the McFadden[41] random utility model (RUM). The modeling approach systematically integrates taste heterogeneity while simultaneously accounting for nonlinear transformations in deprivation cost formulation. This study applies two statistical models: the multinomial logit (MNL) and mixed logit (ML) models. This study explores preference homogeneity and heterogeneity for various attribute specifications, and incorporates mathematical modeling based on Box–Cox and exponential transformations for deprivation time.

This study advances the field through three key contributions: This is the first study that develops a comprehensive DCF framework specifically for infrastructure service disruptions such as electricity, thereby extending beyond traditional applications. Second, while the electricity restoration is immediate, it employs the cost through monthly electricity bills as the price attribute, thus establishing a realistic monetary attribute that increases consequentiality of choices. Third, the methodology offers transferable applications to other infrastructure, such as transportation systems, enabling measurement of social impacts from infrastructure disruptions in economic analysis of resilience investment. By bridging the gap between traditional infrastructure metrics and comprehensive social impact assessment, this framework ultimately enables optimized investment decisions for power systems and transportation networks vulnerable to disruptions. Three findings arise from this study. First, DCFs are convex and strictly increasing with deprivation time, thus confirming theoretical propositions that the user cost of being deprived of electricity is non-linear with respect to time. Second, sociodemographic interaction effects, such as children's presence, also affect the deprivation cost function, thereby revealing how different population segments respond to infrastructure deprivation. The presence of children in households significantly increases willingness to pay for expedited power restoration during outage events. Third, residents in lower-income areas are more sensitive to deprivation time than those in high-income areas.

## 2. Literature

The economic assessment of the impact of disasters upon infrastructure has shifted from a narrow focus on physical infrastructure damage to encompass broader societal effects in recent decades[42,43]. This evolution has employed a comprehensive framework of interrelated concepts—including social cost, willingness-to-pay, utility theory, deprivation of goods and services and the associated deprivation cost, and deprivation cost as a function of time deprived (DCFs)—that captures the multifaceted socioeconomic consequences of disasters. These theoretical constructs not only provide tools for quantifying human suffering in economic terms but also inform allocation strategies for humanitarian resources in post-disaster contexts.

At the heart of the DCF framework lies the concept of social cost, which represents the aggregation of all societal impacts resulting from disasters. Unlike traditional cost accounting that emphasizes direct damage to infrastructure and property, social cost encompasses both private logistics costs and the externalities associated with human suffering caused by the lack of critical supplies and services[44]. These effects, commonly referred to as deprivation costs, capture the economic valuation of individual welfare losses experienced when communities face delays in receiving essential goods, such as food, water, medical care, and electricity, following catastrophic events[20,45]. They represents a crucial advancement in disaster economics by acknowledging that time without essential services imposes measurable welfare losses on affected populations[46,47]. Empirical research has demonstrated that these costs exhibit distinctive mathematical properties: they are typically nonlinear, convex, and monotonically increasing functions of deprivation duration[31,45]. This nonlinearity reflects a fundamental insight about human suffering—each additional hour without critical supplies imposes disproportionately greater hardship than the previous hour, suggesting that marginal disutility increases as deprivation persists.

The recognition of these intangible costs marks a paradigm shift in disaster economics, acknowledging that the true burden of disasters extends far beyond physical destruction to encompass profound human suffering that disproportionately affects vulnerable populations. Underpinning the quantification of these welfare losses is utility theory, which provides the theoretical foundation for understanding how individuals make decisions and experience well-being under various conditions[48,49]. In the context of disasters, utility theory posits that affected individuals seek to maximize their well-being subject to constraints imposed by resource scarcity and service disruptions. When critical infrastructure fails and supply chains collapse, individuals must navigate complex decision-making scenarios where they balance competing needs against limited alternatives. Random utility theory offers a robust framework for modeling preferences and choices under deprivation conditions that characterize post-disaster environments.

The translation of theoretical utility losses into economic metrics relies heavily on the concept of willingness-to-pay, which serves as an estimation of welfare economics[50]. WTP measures the monetary value individuals assign to mitigating or avoiding deprivation, essentially capturing how much they would be willing to spend to prevent or reduce the suffering associated with delayed service delivery[32,51,52].

Methods used to estimate DCFs are prominently contingent valuation and discrete choice experiments. Holguín-Veras[31] relied upon contingent valuation methods to establish initial empirical DCF estimates, highlighting their practical applicability in humanitarian logistics models. Alternatively, discrete choice modeling approaches, such as those proposed by Cantillo[45], allow for detailed consideration of demographic heterogeneity and individual preference variability, enhancing the precision and equity of deprivation cost estimates. Pernett[53] introduced innovative models incorporating random regret minimization heuristics, proposing that individuals under extreme stress may follow decision-making strategies that deviate from classical utility maximization. This behavioral economics perspective suggests that emergency conditions trigger cognitive biases and heuristic-based choices that fundamentally alter how people value resources and make trade-offs.

## 3. Methodology

### 3.1 Research Overview

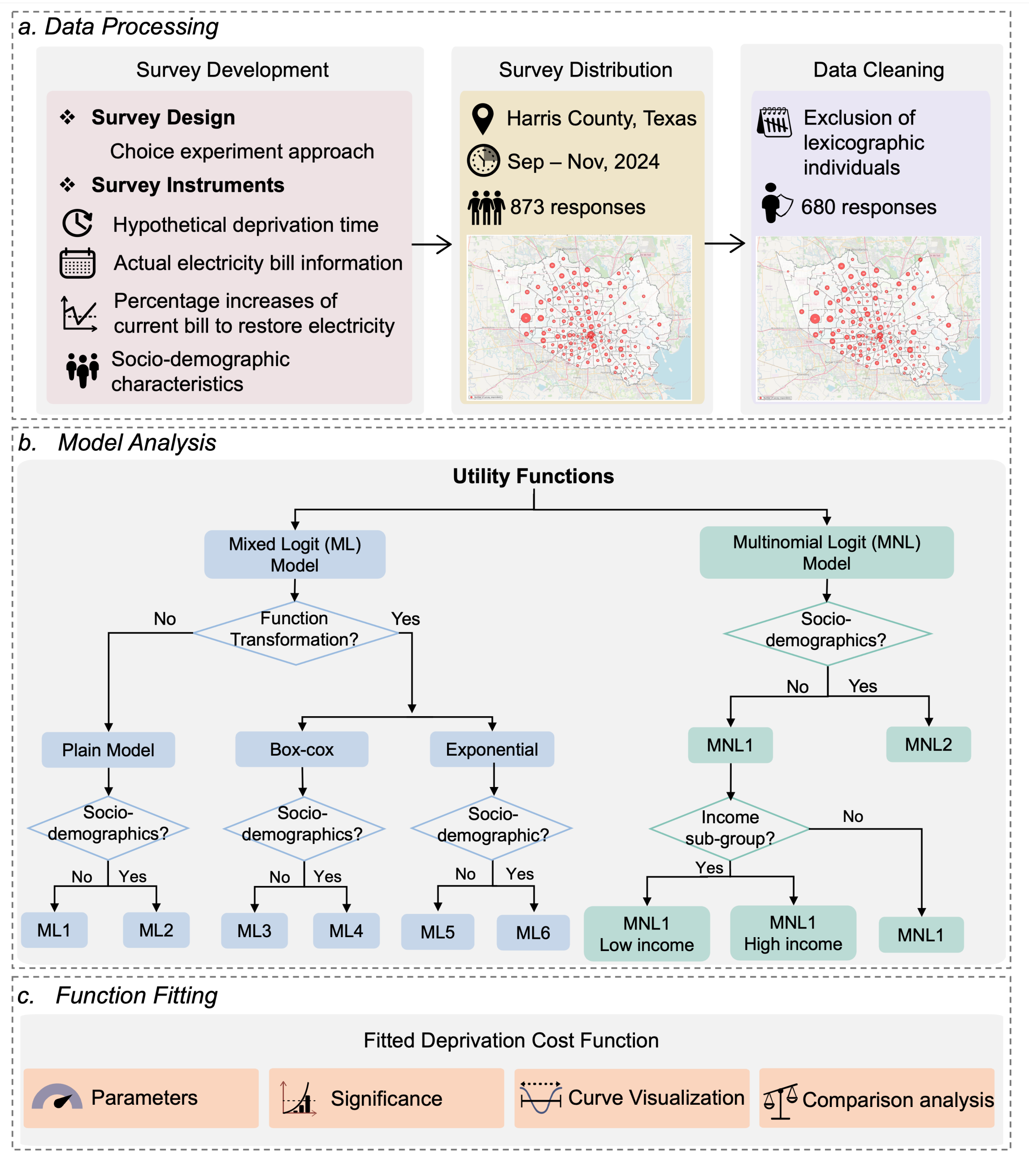

**Figure 1. Overview of the study. This study includes three main steps: a. Data Processing; b. Model Analysis; c. Function Fitting**

Figure 1 presents the methodological framework employed in this study to quantify deprivation costs associated with power outages. The research follows a three-stage process: a. Data Processing; b: Model Analysis; c: Function Fitting. The study area is Harris County, Texas, which has been subject to several severe power outages during natural hazards[3,54]. In step a, the survey was developed to simulate essential variables, including hypothetical deprivation time (including already deprived and expected additional waiting time), actual average electricity bill information, percentage increases of current bill to restore electricity. After pre-test the experiment, the survey was subsequently distributed throughout Harris County, Texas, during the period from September through November 2024, eliciting 873 completed responses. After exclusion of lexicographic individuals, 680 valid responses were used for subsequent analysis. The cleaned dataset serves as the foundation for step b. Model Analysis. In step b, this study developed five distinct families of utility functions: MNL1 and MNL2; MNL1 Low Income and MNL1 High Income; ML1 and ML2; ML3 and ML4; and ML5 and ML6. These models' families differ based on their incorporation of socio-demographic variables, application of functional transformations, or utilization of specific data subsets for estimation purposes. The analysis uses a range of econometric models and mathematical formulations designed to model both homogeneous and heterogeneous preferences. These include sociodemographic variants of multinomial logit models, mixed logit specifications incorporating various functional forms (Box-Cox, and exponential transformations), and models for high-income and low-income population segments. In step c, this study used the significant parameters to estimate the corresponding multiple DCF. The utility estimates derived from these models subsequently inform the construction of a social cost function in form of the deprivation cost functions, which serve as the primary analytical output. These functions characterize the nonlinear relationship between outage duration and welfare losses, providing parameter estimates and statistical significance tests that quantify how deprivation costs accumulate over time.

## 3.2 Modeling Approach

The estimation of deprivation cost functions in this study is predicated on the principles of random utility model[41], a widely accepted framework for analyzing choice behavior. In the RUM framework, individuals are assumed to choose the option that is expected to yield the highest level of utility from among multiple options. Analysts use this framework to quantify preferences and trade-offs individuals make when faced with decisions under constraints, such as when they are asked to choose whether to procure essential supplies or endure further deprivation in a post-disaster context.

### *3.2.1 Random Utility Theory*

The core assumption of RUM is that an individual $n$, when presented with a set of $j$ alternatives, will select the alternative $j$ that yields the maximum perceived utility[55,56]. The utility $U_{nj}$ can be modeled as a deterministic (or systematic) component, $V_{nj}$, and a stochastic (or random) component, $\epsilon_{nj}$ :

$$U_{nj} = V_{nj} + \epsilon_{nj}. \tag{1}$$

The systematic component, $V_{nj}$, is a function of the observable attributes of the alternative $X_{nj}$, and the characteristics of the individuals $S_n$ [56,57]. The random component, $\epsilon_{nj}$, accounts for all unobserved influences on choice, including unmeasured attributes, variations in tastes, and measurement errors by the analyst.
The probabilistic nature of choice is captured by assuming a specific distribution for the random error terms $\epsilon_{nj}$. If these error terms are assumed to be independently and identically distributed (IID) following a Gumbel distribution, the probability $P_{nj}$ that individual *n* chooses alternative *i* from a choice set $A_{(n)}$ is given by the multinomial logit (MNL) model[41]:

$$P_{nj} = \frac{e^{(V_{nj})}}{\sum_{Ai \in A(n)} e^{(V_{ni})}}. \quad (2)$$

The systematic utility $V_{nj}$ is commonly specified as a linear-in-parameters function for analytical tractability[58]. A general form is:

$$V_{nj} = ASC_j + \sum_k \beta_{njk} \chi_{njk}, \quad (3)$$

where $ASC_j$ represents the alternative-specific constant for alternative *j,* which captures the average influence of all factors not explicitly included in the model for that alternative, effectively representing the baseline utility of alternative j when all $\chi_{njk}$ are zero. The term $\chi_{njk}$ denotes the $k^{th}$ attribute of alternative $j$ as faced by individual $n$ (e.g., cost of the supply, deprivation time endured). The parameter $\beta_{njk}$ is the marginal utility or taste weight associated with attribute $\chi_{njk}$ for individual *n*. A positive $\beta_{njk}$ implies that an increase in $\chi_{njk}$ increases utility, while a negative $\beta_{njk}$ implies the opposite. When $\beta_{njk}$=$\beta_{jk}$, preferences are identical among all individuals, and it corresponds to the MNL model.

However, individuals often exhibit varied preferences. This heterogeneity can be addressed in several ways. To account for preference variations that can be explained by observable characteristics of the individuals (e.g., income, age, household composition), the taste parameters $\beta_k$ can be specified as a function of these socioeconomic characteristics $S_{nl}$ :

$$\beta_{njk} = \beta_{jk} + \sum \delta_{jkl} S_{nl}. \quad (4)$$

Substituting Equation (4) into Equation (2) allows the systematic utility to reflect these observed sources of heterogeneity:

$$V_{nj} = ASC_j + \sum_k (\beta_{jk} + \sum_1 \delta_{jkl} S_{nl})\, x_{njk}. \quad (5)$$

The parameters $\delta_{jkl}$ capture how the marginal utility of attribute $k$ changes with the socioeconomic characteristic $l$.

Beyond observed characteristics, preferences can vary randomly across the population due to unobserved factors. Mixed logit (ML) models[56] accommodate this by allowing taste parameters $\beta_{nk}$ to be random variables, following a specified distribution $f(\beta|\theta)$ with parameters $\theta$ (e.g., mean $\bar{\beta}_k$ and standard deviation $\sigma_k$ if normally distributed). The utility function for an ML model can be expressed by incorporating an individual-specific deviation $\eta_{nk}$ (drawn from the specified distribution) from the population mean parameter $\bar{\beta}_k$:

$$V_{nj} = ASC_j + \sum_k (\beta_{jk} + \sum_1 \delta_{jkl} S_{nl} + \eta_{jkn}) f(x_{njk}). \quad (6)$$

[1]

*3.2.2 Functional Form for Attributes on Deprivation Time*

A critical attribute in this research is deprivation time. Theoretical considerations suggest that the disutility (or cost) associated with deprivation is not linear but rather increases at an accelerating rate, implying a convex DCF.[30] To capture such non-linear effects within the linear-in-parameters utility framework, the DT attribute is transformed. Common functional transformations include Box-Cox transformation and exponential transformation. The Box-Cox transformation is flexible and allows the data to determine the most appropriate functional form (Box & Cox, 1964). If $x > 0$ is the time that the individual is deprived of electricity, then the resulting disutility can be specified as proportional to the Box-Cox transformation:

$$x^{(\tau)} = \begin{cases} \frac{(x^{\tau}-1)}{\tau}, & if\ \tau \neq 0 \\ \ln(x), & if\ \tau = 0 \end{cases}, \quad (7)$$

where the transformation parameter $\tau$ is estimated alongside other utility parameters. If $\tau = 1$, the relationship is linear. If $\tau<1$, $x^{(\tau)}$ is concave in $x$, while if $\tau > 1$ the function captures increasing marginal disutility (convexity in cost terms). $x$ in this case is time, and $x^{(\tau)}$ is a non-variable that affects utility, thus $\beta \cdot x^{(\tau)}$ is the effect of this variable on utility.

*3.2.3 Derivation of Deprivation Cost Functions*

The estimated utility functions serve as the basis for deriving DCFs. DCFs represent the monetary valuation of the welfare loss (or suffering) due to deprivation. This is typically achieved by calculating the willingness to pay (WTP) to avoid deprivation or, equivalently, the change in consumer surplus ($CS_n$) resulting from an increase in deprivation time[45]. The valuation of utility changes relies on estimates of consumer surplus variations, which assumes individuals make choices through compensatory decision-making mechanisms[59]. Therefore, the willingness-to-pay measure is determined by calculating the marginal rate of substitution between the attribute (deprivation time) and the monetary attribute (Income). The standard marginal value of deprivation time $MVDT_n$ defines how willing is the individual to pay for saving deprivation time:

$$MVDT_n = \frac{\partial V_n / \partial T_n}{\partial V_n / \partial I_n}. \tag{8}$$

---

[1] $f(x_{njk})$ is used here to also denote that attributes might be transformed, as discussed next.)

The marginal utility of income, $\partial V_n / \partial I_n$, is the negative of the cost coefficient, which means $\partial V_n / \partial I_n = -\beta_c$[56].

The change in consumer surplus then can be calculated by the production of $MVDT_n$ and the deprivation time:

$$\Delta E(CS_n) = (MVDT_n)\,(T^0 - T^1). \tag{9}$$

However, If the utility function is non-linear with respect to *DT* (due to transformations like Box-Cox or exponential), then $\partial V_n / \partial T_n$ will be a function of *DT*, making $MVDT_n$ also a function of *DT*. The total deprivation cost for an increase in deprivation from an initial time $DT_0$ to a final time $DT_1$ is then the integral of the $MVDT_n$ over this period:

$$\Delta E(CS_n) = \int_{T^0}^{T^1} MVDT_n(t)\,dt. \tag{10}$$

For ML models, $MVDT_n$ is computed for each draw of the random parameters, the summation or integration is performed, and the resulting DC values are averaged across draws. Graphically, a typical DCF curve represents the relationship between deprivation costs (y-axis) and deprivation time (x-axis). Initially, the curve rises slowly, representing lower levels of discomfort or suffering at shorter deprivation durations. As deprivation time increases, the slope of the curve steepens significantly, reflecting escalating levels of human suffering and a heightened urgency for relief supplies. Eventually, the curve approaches a terminal point representing the maximum deprivation threshold, beyond which survival becomes increasingly uncertain, thus capturing the ultimate limit of human endurance in the absence of critical goods or services.

## 4. Case Study

### 4.1 Data

This study utilized data collected from a sample of individuals from Harris County (Texas), conducted between September and December 2024, between two and five months after Hurricane Beryl. The sample is drawn from the Qualtrics panel, members of which are recruited to be representative of the underlying population. The collected data [2] encompassed two primary components: (1) comprehensive socioeconomic information at both household and individual levels, including demographic factors (age, gender, household size, number of children), economic indicators (income level, occupation), educational attainment, and experience with recent events (Hurricane Beryl, May storm, and previous power outages); and (2) respondent choice preferences derived from the stated preference using choice experiment method.

The experiment was designed to simulate infrastructure deprivation scenarios in which respondents' homes experienced extended power outages. The choice experiment is based on an efficient design with 36 total choice scenarios divided into 9 blocks of 4 questions each. The efficient design is generated using Ngene® software and is based on prior information from the pretest phase to ensure precise estimation of preferences with a minimal number of choice tasks.[3] In selected choice scenarios, respondents were presented with the option to pay an additional cost on their electricity bill to immediately restore power service or wait more time for the restoration of power. Prior to presenting the four choice scenarios, respondents were asked to report their current average monthly electricity bill in dollars (B). This value was used to pivot the cost attribute in the choice experiment relative to each respondent's baseline, thereby increasing the saliency and realism of the cost variations in the design. A respondent whose average monthly bill is $100 would be presented with the following scenario:

> "*Imagine a major weather event like a hurricane or tornado has knocked out power in your area, leaving you and your family without electricity. There is no set time for when power will be back on, and based on past events, while on average, power outages are less than 2 hours, some Texans have experienced power outages that have lasted up to two weeks. You now face a choice about how to deal with this blackout. You can opt to have a backup power system (such as a generator) that immediately restores electricity specifically to your home. The cost of this service will be added to your electricity bill and paid off over the next 12 months. Alternatively, you can wait and see when the power will be restored naturally.*
>
> *"The options will vary in several ways. Select which option you prefer in each scenario. The following are options that you will see in choice scenarios:*

[2] The IRB number is STUDY2024-0699

[3] The D-error of the selected efficient design is 0.0058.

- *Already deprived time (Time that you have already been deprived of electricity): 1, 3, 5, 7 (days)*
- *Additional time (Additional time that you would be without electricity): 1, 3, 5, 7 (days)*
- *Electricity bill percentage increase over the next 12 months: 10%, 25%, 50%, 75% and 100% of current monthly payment*

  *Although these choices are hypothetical, please respond as if you were actually deciding whether to accept the purchase offer or wait for electricity to be restored naturally. Each scenario is unique, so it's important to consider each one carefully and respond to all 4 scenarios.*

  *Scenario #1 of 4:*

  *Suppose that you have already been without electricity for 3 days and the electricity company has stated that you need to wait 1 more day before your electricity is restored. Would you pay $150} (a 50% increase in your current electricity bill) per month over the next 12 months to restore your electricity immediately?*

  *O Yes, I would pay $150} per month over the next 12 months.*

  *O No, I would wait one more day without electricity and see no increase in my electricity bill."*

The choice experiment incorporated three key attributes shown in Table 1: (i) current deprivation time, representing the duration without electricity since the event occurred; (ii) waiting time (WT), indicating the additional time required before power restoration would occur naturally; and (iii) the monthly electricity bill after a percentage increase if they elected to purchase the backup power system.

**Table 1**
**Variables in experimental design.**

| **Variables** | **Values** |
|---|---|
| *DT:* Current deprivation time (days) | *1, 3, 5, 7* |
| *WT:* Additional waiting time (days) | *1, 3, 5, 7* |
| *B*: Current electricity bill | Respondent's monthly electricity bill |
| *P*: Increasing percentage | 10%, 25%, 50%, 75% |

Each respondent was randomly assigned to one of 9 blocks and presented with four choice situations. In each case, respondents were presented with two options: purchase an immediate backup power solution at an increased cost (percentage increase of current electricity bill) or maintain their current electricity bill and wait for natural power restoration after WT. The purchasing option was directly associated with the current deprivation time, representing how long respondents had already been without power. Conversely, the waiting option was associated with an expected total deprivation time (EDT), calculated as the sum of current deprivation time and additional waiting time (EDT = DT + WT). To mitigate potential order bias, the presentation sequence of scenarios was randomized across participants. This design methodology ensured attribute level balance within each block.

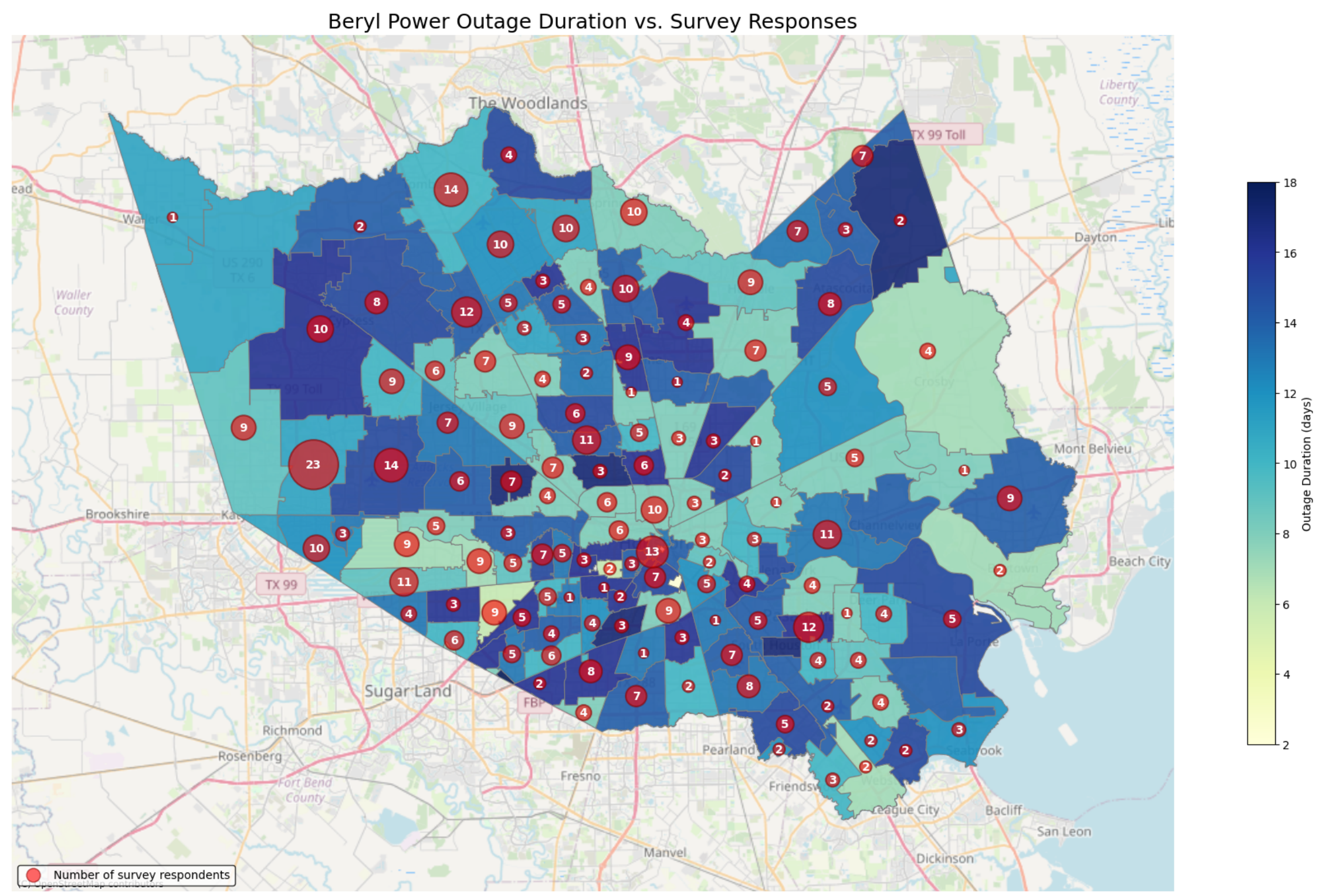

**Figure 2. Survey respondents' distribution and Beryl power outage duration across ZCTA areas in Harris County, Texas.**

Figure 2 displays the spatial distribution of survey respondents across ZIP Code Tabulation Areas (ZCTAs) and example of Beryl power outage within Harris County, as Beryl power outage is the most recent significant power outage event. Survey data from this study also reveals that more than 90% of respondents encountered power disruptions during either the May 2024 storm, Hurricane Beryl in 2024, or during both events. The visualization employs a dual-representation approach where the ZCTA area color gradient indicates outage duration measured in days, ranging from approximately 2 days (shown in light green) to 18 days (displayed in dark blue). The red circle positioned within each ZCTA represents the number of survey responses collected from that geographic area, with circle size corresponding to response volume. Qualtrics implemented comprehensive data collection strategies to ensure representative coverage across most ZIP Code Tabulation Areas within Harris County, recognizing the critical importance of capturing diverse experiences throughout the region. The study initially collected a total of 873 responses, which were subsequently processed, coded, and reviewed to ensure data quality and integrity. After excluding lexicographic responses and conducting data depuration, 680 surveys remained for modeling purposes.

The final sample represented diverse socioeconomic conditions. Respondents ranged in age from 18 to 93 years, with an average age of 47 years. The gender distribution showed a slight female majority (54.9%) compared to male respondents (45.1%). Regarding household composition, 46.6% of respondents reported family sizes of 1–2 members, 36.6% had households of 3–4 members, and 13.7% had larger families of 5–6 members. Additionally, 43.2% of respondents reported having at least one child in their house.

When questioned about disaster experience, most respondents (51.6%) were affected by both Hurricane Beryl and a serious thunderstorm in May, which also resulted in power outages; 36.2% were affected only by just Hurricane Beryl; 7.2% were affected only by the May thunderstorm; leaving only 5% who were affected by neither storm. This heterogeneous sample composition enhanced model quality by ensuring independent variables covered a comprehensive range of values, which explains the deliberate inclusion of diverse communities and population groups with varying disaster experiences. Table 2 provides a more detailed breakdown of the sample's socioeconomic characteristics.

Analysis of responses to the hypothetical choice scenarios revealed notable patterns related to household composition. Specifically, households without children demonstrated a stronger preference for the "wait" option (77.8%) compared to the "purchase" option (22.2%). In contrast, households with one or more children showed a higher propensity to select "purchase" (31.2%) and a correspondingly lower tendency to choose "wait" (68.8%) when compared to childless households. The data collected through the stated choice experiment were subsequently used to estimate the multinomial logit and mixed logit models using the simulated maximum likelihood method. The detailed modeling approach and results are presented in the following section.

**Table 2**
**Sample statistics**

| Variable | Number | Percentage | Variable | Number (n) | Percentage (%) |
|---|---|---|---|---|---|
| **Gender** | | | **Ethnicity** | | |

| Male | 307 | 45.1% | White | 246 | 36.2% |
|---|---|---|---|---|---|
| Female | 373 | 54.9% | Hispanic or Latino | 212 | 31.2% |
| **Age** | | | Black or African American | 192 | 28.2% |
| 18 - 25 | 99 | 14.6% | Asian/Pacific Islander | 9 | 1.3% |
| 26 - 40 | 188 | 27.6% | Native American or American Indian | 5 | 0.7% |
| 41 - 65 | 265 | 39.1% | Others | 8 | 1.2% |
| > 65 | 127 | 18.7% | Prefer not to answer | 8 | 1.2% |
| **Household Size** | | | **Annual Income** | | |
| 1 - 2 | 317 | 46.6% | < \$25,000 | 103 | 15.1% |
| 3 - 4 | 249 | 36.6% | \$25,000 - \$49,999 | 144 | 21.2% |
| 5 - 6 | 93 | 13.7% | \$50,000 - \$74,999 | 171 | 25.1% |
| 7 - 9 | 20 | 2.9% | \$75,000 - \$99,999 | 78 | 11.5% |
| > 10 | 1 | 0.1% | \$100,000 - \$149,999 | 97 | 14.3% |
| **Children Number** | | | > \$149,999 | 73 | 10.7% |
| 0 | 386 | 56.8% | Prefer not to answer | 8 | 1.2% |
| 1 | 135 | 19.9% | **Storm Experience** | | |
| 2 | 96 | 14.1% | Hurricane Beryl | 246 | 36.2 |
| 3 | 29 | 4.3% | May Thunderstorm | 49 | 7.2% |
| 4 | 21 | 3.1% | Both | 351 | 51.6% |

### 4.2 Modeling framework

To analyze how deprived time in conjunction with socioeconomic and environmental constraints affect decision-making during power outages and to evaluate their impact on deriving accurate deprivation cost functions, we developed a set of multinomial logit and mixed logit (ML) models. These models are specifically designed for choice analysis among finite alternatives, aligning with our stated choice experiment. The mixed logit framework is particularly advantageous for incorporating individual-level variations and repeated measures, essential for understanding nuanced responses to power outage conditions. Table 3 shows the attributes used in this study for the utility function, including current deprivation time, additional waiting time, final monthly electricity bill and children binary variable.

**Table 3**

Attributes used in the specification of utility functions

| Notation | Variables | Description |
|---|---|---|
| DT | Current deprivation time (days) | |
| WT | Additional waiting time (days) | |
| C | Final monthly electricity bill | Current electricity bill *(1+P) |
| CH | Binary variable associated with the presence of children at home | 1: If children number > 20% percent of household number; 0: Otherwise |

Table 4 specifies four families of MNL and ML models with capabilities of considering the nonlinear behavior of DCFs.

The notation in Table 4:

$U_{n(j)}$ Utility of alternative $j \in \{purchase\ (p), wait\ (w)\}$ associate to individual n

ASC: Constant for alternative purchase ($p$)

$\beta_c$ Parameter associated to C

$\beta_t$ Parameter associated to DT or DT+WT

$\delta_{CHT}$ Parameter associated to the interaction between high presence of children and deprivation time

$\tau$ Parameter of the Box-Cox function

$\beta_T$ Parameter related to deprivation time, only for the exponential model

The $\epsilon$ is the Gumbel IID distribution. The additional error component $\xi \sim Normal\ (0, \sigma_\xi^2)$ is to consider the panel effect in all alternatives.

In Table4, the first family, MNL1 and MNL2, utilized MNL models to estimate specifications incorporating current deprivation time (DT), additional wait Time (WT), final cost (C). MNL2 included a binary variable for children (CH) as observable attributes or sociodemographic variables. MNL1 low-income, and MNL1 high-income were used same way as MNL1, but we divided the data into high income and low income based on their income level. The ML1 and ML2 model families utilize mixed logit specifications that incorporate DT, WT, and C. These models leverage the panel structure of the data to capture individual heterogeneity, with ML2 additionally including interactions with the binary children variable. To more explicitly address nonlinearities in time-related attributes, the ML3 and ML4 family specified models with Box-Cox transformations of time parameter. Within this family, both incorporates the panel effect, while ML4 further incorporated systematic heterogeneity by interacting time attributes with the children binary variable. Similarly, the ML5 and ML6 family utilized exponential transformations for time-related attributes, with ML6 including terms for observed heterogeneity based on socioeconomic interactions. All specified models in Table 4 were estimated using simulated maximum likelihood techniques. This study does not incorporate random heterogeneity in preferences toward deprivation time but to the ASC, as preliminary testing indicated this parameter was not statistically significant in the model specification.

**Table 4**
**Specifications of the models**

**MNL1; MNL1 low-income; MNL1 high-income**

$$U_{np} = ASC + \beta_c \cdot C_{nt} + \beta_t \cdot DT + \epsilon_{ntp}$$
$$U_{nw} = \beta_t \cdot (DT + WT) + \epsilon_{ntw}$$

**MNL2**

$$U_{np} = ASC + \beta_c \cdot C_{nt} + (\beta_t + \delta_{CHT} \cdot CH) \cdot DT + \epsilon_{ntp}$$
$$U_{nw} = (\beta_t + \delta_{CHT} \cdot CH) \cdot (DT + WT) + \epsilon_{ntw}$$

**ML1**

$$U_{np} = ASC + \beta_c \cdot C_{nt} + \beta_t \cdot DT + \xi_{np} + \epsilon_{ntp}$$
$$U_{nw} = \beta_t \cdot (DT + WT) + \epsilon_{ntw}$$

**ML2**

$$U_{np} = ASC + \beta_c \cdot C_{nt} + (\beta_t + \delta_{CHT} \cdot CH) \cdot DT + \xi_{np} + \epsilon_{ntp}$$
$$U_{nw} = (\beta_t + \delta_{CHT} \cdot CH) \cdot (DT + WT) + \epsilon_{ntw}$$

**ML3**

$$U_{np} = ASC + \beta_c \cdot C_{nt} + \beta_t \cdot DT^{\tau} + \xi_{np} + \epsilon_{ntp}$$
$$U_{nw} = \beta_t \cdot (DT + WT)^{\tau} + \epsilon_{ntw}$$

**ML4**

$$U_{np} = ASC + \beta_c \cdot C_{nt} + (\beta_t + \delta_{CHT} \cdot CH) \cdot DT^{\tau} + \xi_{np} + \epsilon_{ntp}$$
$$U_{nw} = (\beta_t + \delta_{CHT} \cdot CH) \cdot (DT + WT)^{\tau} + \epsilon_{ntw}$$

**ML5**

$$U_{np} = ASC + \beta_c \cdot C_{nt} + \beta_t \cdot e^{\beta_T \cdot DT} + \xi_{np} + \epsilon_{ntp}$$
$$U_{nw} = \beta_t \cdot e^{\beta_T \cdot (DT+WT)} + \epsilon_{ntw}$$

**ML6**

$$U_{np} = ASC + \beta_c \cdot C_{nt} + (\beta_t + \delta_{CHT} \cdot CH) \cdot e^{\beta_T \cdot DT} + \xi_{np} + \epsilon_{ntp}$$
$$U_{nw} = (\beta_t + \delta_{CHT} \cdot CH) \cdot e^{\beta_T \cdot (DT+WT)} + \epsilon_{ntw}$$

**Results and analysis**

The estimation results for multinomial logit and mixed logit models utilizing both Box-Cox and exponential specifications are summarized in Table 5. Except for the specifications that divide the sample into income groups, the results are based on 2,720 observations corresponding to four choice scenarios collected from 680 individuals. Most of the parameters are statistically significant at least at the 95% confidence level, and the signs are consistent with economic theory expectations.

The econometric modeling efforts yielded substantial insights into the economic valuation of deprivation. The examination of all estimated models, encompassing multinomial logit, various mixed logit specifications including those with linear or polynomial time effects, and ML models incorporating specific transformations, such as Box–Cox and exponential for the time attribute, reveals that the core estimated parameters are robust and consistent with established economic theory and the relevant research within humanitarian logistics. The cost parameter, for instance, was consistently negative and demonstrated statistical significance across all models, thereby indicating that an increase in the monetary outlay required to procure essential goods corresponds with a diminished likelihood of their selection. In a similar vein, the deprivation time parameter consistently exerted a significant negative influence on utility. This underscores the principle that extended periods lacking essential supplies lead to a substantial escalation in disutility, or human suffering. This critical finding lends further empirical support to the central tenets of deprivation cost theory.

Notably, the Box–Cox parameter estimates, approximately 1.27–1.28 in ML3 and ML4, confirm the nonlinear nature of the deprivation cost relationship with deprivation time.

Similarly, the exponential time parameters are positive (0.0464 and 0.0501 for ML5 and ML6, respectively) and significant, reinforcing the strictly convex relationship between DC and DT. Significant standard deviations for panel effects in ML models suggest considerable random heterogeneity across respondents. The parameter associated with the interaction between the presence of children and deprivation time is statistically significant in models MNL2 and ML2. Although it is marginally significant at the 90% confidence level in models ML4 and ML6, the negative sign of these parameters underscores a heightened sensitivity to infrastructure deprivation among households with children. Moreover, the significant standard deviation associated with the panel effect in all ML models highlights the existence of substantial unobserved heterogeneity among individuals in their valuation of deprivation. This observation validates the use of ML specifications, as opposed to simpler MNL models, to adequately account for the random variations in taste and the inherent correlation of choices made by the same respondent over multiple scenarios.

Income-based segmentation provides additional insights. The coefficient for low-income (-0.1969) and high-income (-0.1605) groups reveals differential sensitivity towards deprivation. Interestingly, lower-income households display a higher sensitivity to deprivation costs, indicative of greater valuation placed on timely infrastructure restoration. This heightened sensitivity likely stems from the reduced resilience capacity of lower-income communities, which possess fewer resources to mitigate the negative impacts of power outages[3,60,61]. Low socioeconomic status individuals have limited capacity to store food and water or own a generator, constraining their ability to maintain essential services during extended outages.

Figs. 3–7 illustrate fitted deprivation cost functions derived from utility functions based on welfare estimation presented in equation (11). The resulting curve functions were then generated using polynomial regression. Due to the higher degree of convexity observed in the ML5 and ML6 models, third-degree polynomials were used for these two cases, while quadratic polynomials were applied to the other models. All fits achieved adjusted R-squared values of 0.99, indicating excellent model performance. The graphical representations of the derived DCFs provide visual confirmation of the theoretical expectation that deprivation costs exhibit a strictly increasing and convex relationship with respect to deprivation time. Models that incorporated the children-time interaction (e.g., MNL2, ML2, ML4, ML6) consistently generated higher DCFs than their respective baseline counterparts (MNL1, ML1, ML3, ML5). This suggests a greater societal cost when vulnerable demographic groups are impacted, which is the high presence of children in household in this study. Such an observation lends support to the notion that more comprehensive modeling approaches, which explicitly account for heterogeneity, are likely to yield higher, and arguably more realistic, estimates of welfare loss. The magnitude of the estimated deprivation costs is considerable, frequently ascending to tens of thousands

of dollars over a 30-day period. This valuation aligns with the substantial figures reported in comparable studies and serves to emphasize that these social costs significantly surpass typical logistical or market-based expenditures. In the present study's results, both the Box–Cox ($\lambda \approx 1.27$, significantly greater than 1, indicating convexity) and exponential transformations for deprivation time within the ML framework led to DCFs that were visibly more convex and resulted in higher estimated deprivation costs at longer durations than what would be implied by a simple linear treatment of time. Specifically, observing the derived DCF curves, the Box–Cox specification (models ML3/ML4) demonstrated a pronounced and accelerating increase in deprivation costs. The exponential specification (models ML5/ML6) also captured this escalating trend, with the visual evidence from the fitted polynomial equations on the graphs suggesting that both these transformed approaches lead to significantly higher cost estimations than simpler forms, accurately reflecting the intensifying nature of deprivation. Box–Cox and exponential specifications can sometimes yield different degrees of convexity; in this instance, both affirm the necessity of non-linear treatment for deprivation time to avoid underestimating welfare loss. The cubic functional forms, as suggested by the polynomial equations fitted to the DCFs from models ML5 and ML6 (representing another form of ML with a higher-order polynomial for time), further illustrated the capacity for rapid cost escalation, particularly at the outer ranges of deprivation time.

A particularly noteworthy and central result was identified in the comparative analysis of DCFs for low-income versus high-income cohorts. The models unequivocally indicate a markedly higher DCF for the low-income segment, signifying a greater welfare loss for this group under similar deprivation conditions. This implies that low-income populations are more sensitive to the effects of deprivation. While the prevailing literature often recommends the adoption of generic DCFs to promote equity by disaggregating intrinsic suffering from an individual's ability to pay, the current finding strongly suggests a heightened vulnerability or an elevated marginal utility of the deprived goods for low-income individuals. This, in turn, leads to a more severe impact and a higher socio-economic cost of deprivation upon this group. This pronounced sensitivity of low-income populations carries significant policy implications for the prioritization of aid, consistent with broader discussions on equity within the domain of humanitarian response.

The present modeling endeavor demonstrates the attainment of robust parameter estimates and the derivation of DCFs that are congruent with established theoretical frameworks. The findings suggest the profound economic impact of deprivation, the amplified costs associated with vulnerable populations, and the critical importance of utilizing advanced modeling techniques capable of accommodating heterogeneity and non-linear temporal effects. The significant divergence in perceived deprivation costs between distinct income groups, with low-income populations exhibiting greater sensitivity and higher losses, warrants further scholarly investigation and careful consideration in the equitable formulation and execution of humanitarian relief strategies. This study also conducted comprehensive testing of interaction effects with additional sociodemographic variables, including gender, age, household size, and power outage experience characteristics. The statistical analysis revealed no significant impact from these demographic interactions on preference heterogeneity for deprivation time. Consequently, these non-significant variables are excluded from the presented analysis to maintain model parsimony and focus on statistically meaningful results.

**Table 5**
**Modeling results**

| Parameter (Notation) | Mean (t-test) | | | | | | | | | |
|---|---|---|---|---|---|---|---|---|---|---|
| | MNL1 | MNL2 | ML1 | ML2 | ML3 | ML4 | ML5 | ML6 | MNL1 low-income | MNL1 high-income |
| | | | | | Box-cox | Box-cox | Exponential | Exponential | | |
| Constant (ASC) | -1.0031<br>(-6.87) | -0.9638<br>(-6.52) | -1.3630<br>(-4.98) | -1.3982<br>(-5.19) | -1.3854<br>(-5.17) | -1.3599<br>(-5.02) | -1.3357<br>(-4.94) | -1.3143<br>(-4.92) | -1.2295<br>(-6.51) | -0.6166<br>(-2.50) |
| Cost | -0.0025<br>(-5.45) | -0.0027<br>(-5.60) | -0.0042<br>(-4.46) | -0.0040<br>(-4.56) | -0.0039<br>(-4.51) | -0.0039<br>(-4.49) | -0.0040<br>(-4.44) | -0.0039<br>(-4.49) | -0.0022<br>(-3.68) | -0.0030<br>(-3.88) |
| Deprivation Time | -0.1826<br>(-9.08) | -0.1425<br>(-6.33) | -0.2653<br>(-8.43) | -0.2273<br>(-6.25) | -0.1493<br>(-2.77) | -0.1285<br>(-2.66) | -3.7373<br>(-2.16) | -2.8366<br>(-2.68) | -0.1969<br>(-7.52) | -0.1605<br>(-5.03) |
| Box-Cox parameter | | | | | 1.2756<br>(8.56) | 1.2707<br>(8.60) | | | | |
| Exponential time | | | | | | | 0.0464<br>(3.30) | 0.0501<br>(4.56) | | |
| Children-time | | -0.0901<br>(-4.40) | | -0.0831<br>(-2.15) | | -0.0477<br>(-1.79) | | -1.0510<br>(-1.79) | | |
| Standard deviation, panel effect | | | 1.7445<br>(17.14) | 1.7059<br>(16.28) | 1.7530<br>(16.90) | 1.7172<br>(16.31) | 1.7488<br>(17.09) | 1.7136<br>(16.30) | | |
| Report of the estimation process | | | | | | | | | | |
| Number of observations | 2720 | 2720 | 2720 | 2720 | 2720 | 2720 | 2720 | 2720 | 1672 | 992 |
| Number of respondents | 680 | 680 | 680 | 680 | 680 | 680 | 680 | 680 | 418 | 248 |

| Measurement of adjustment | | | | | | | | | | |
|---|---|---|---|---|---|---|---|---|---|---|
| Final log-likelihood | -1494.71 | -1485.10 | -1371.70 | -1370.05 | -1370.61 | -1368.10 | -1370.71 | -1368.98 | -891.63 | -570.50 |
| Adjusted rho square | 0.271 | 0.272 | 0.200 | 0.198 | 0.207 | 0.207 | 0.205 | 0.208 | 0.261 | 0.313 |

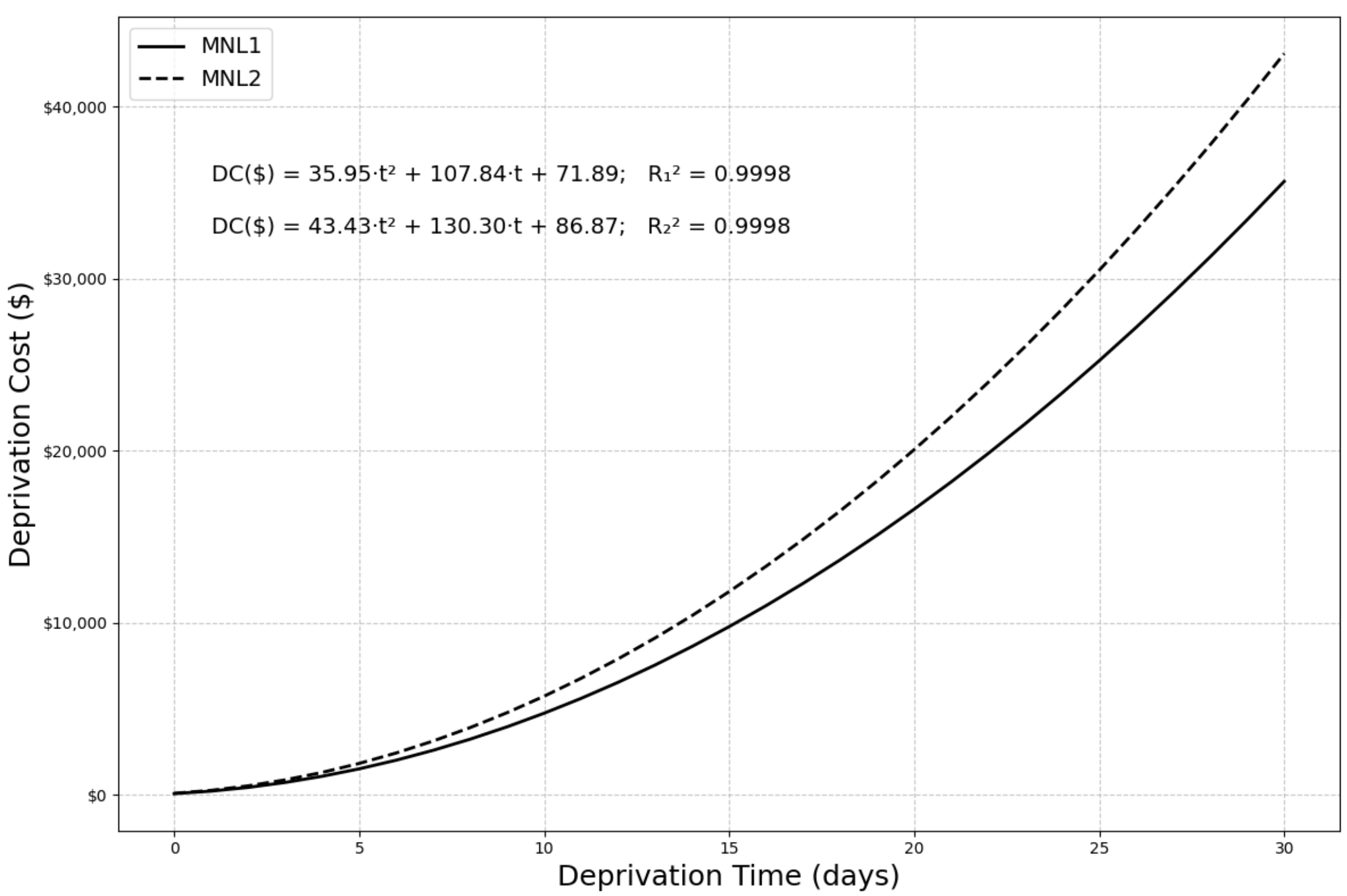

**Figure 3. DCF of MNL1 and MNL2 models.**

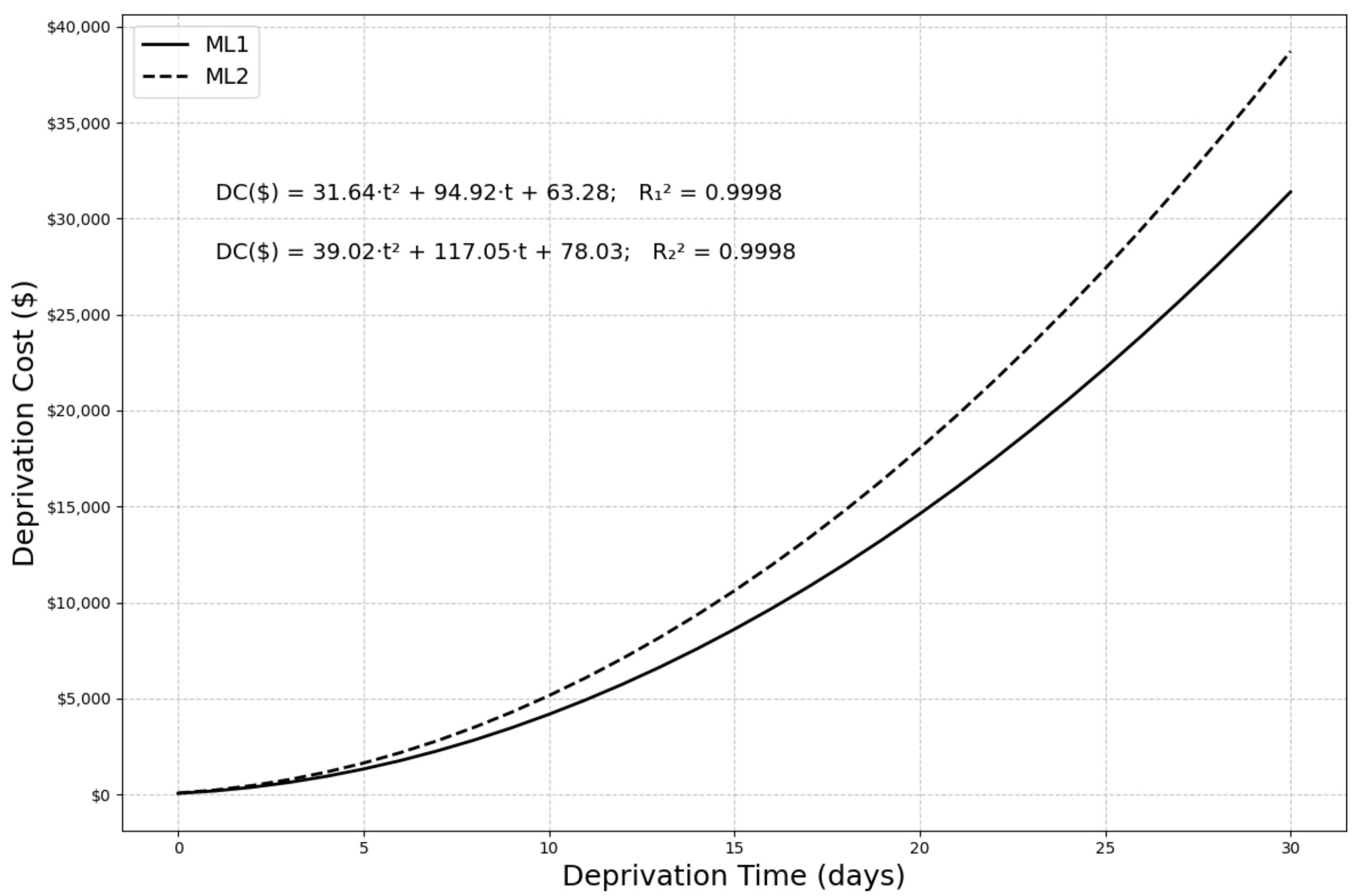

**Figure 4. DCF of ML1 and ML2 models.**

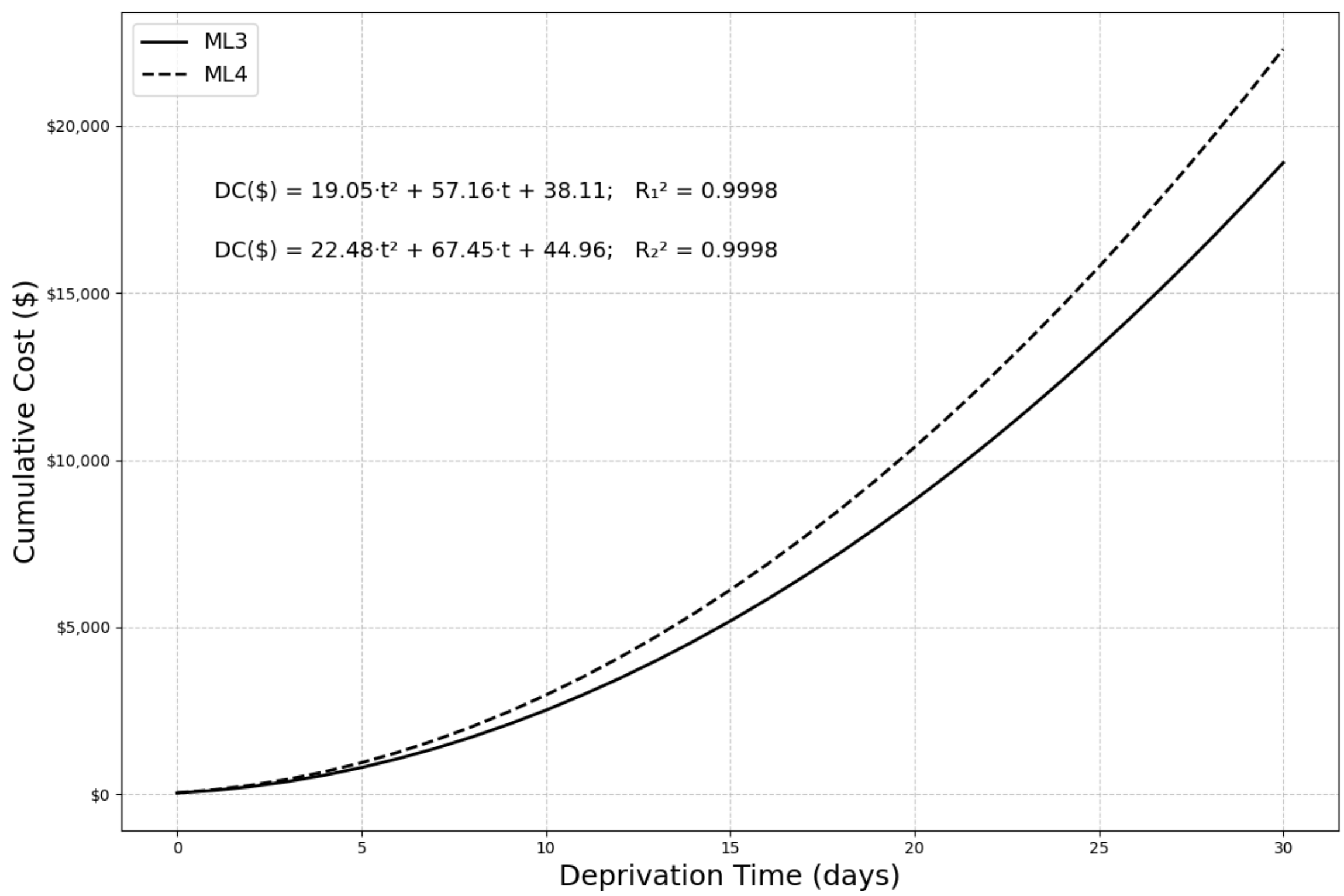

**Figure 5. DCF of ML3 and ML4 models.**

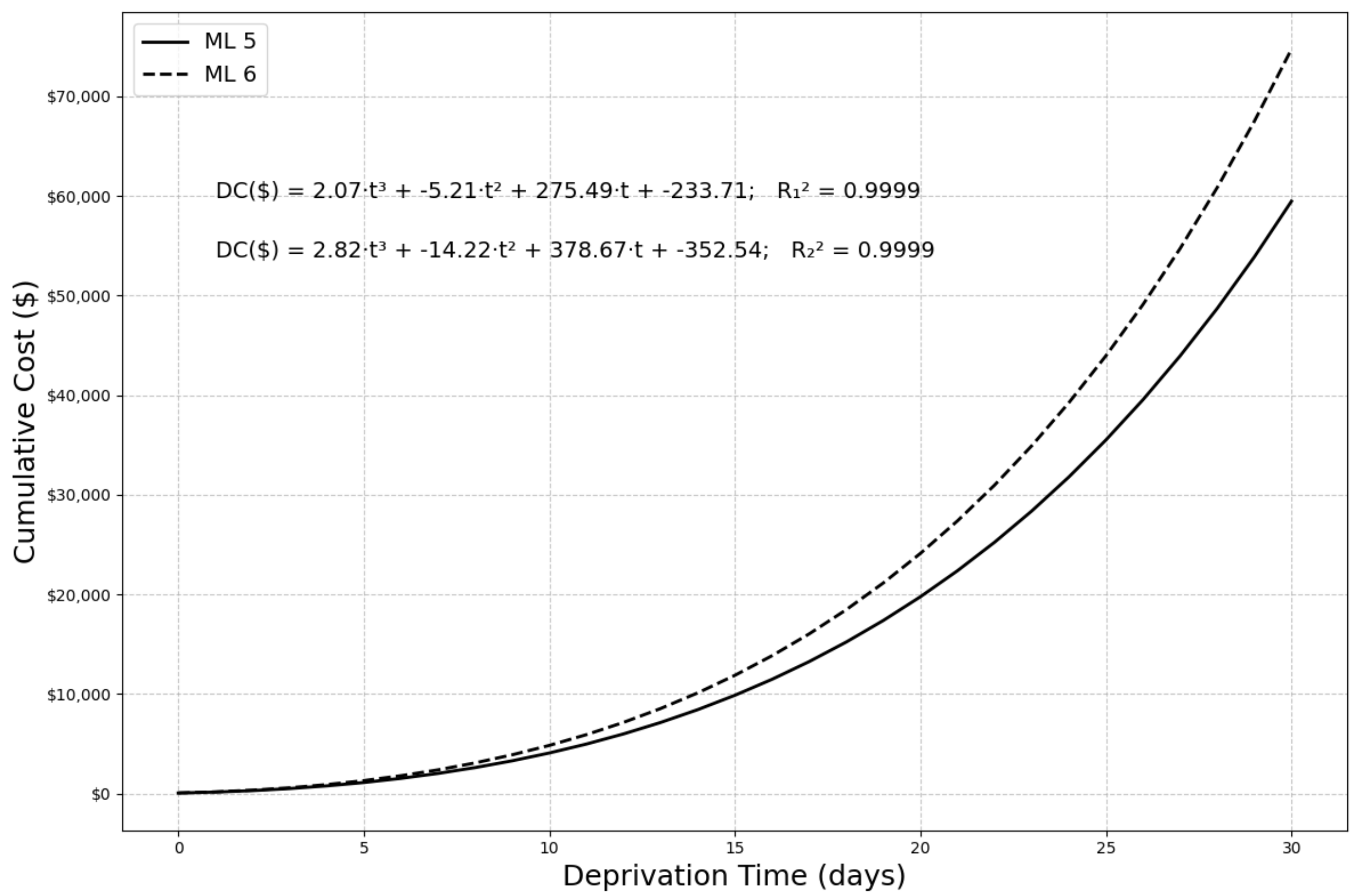

**Figure 6. DCF of ML5 and ML6 models.**

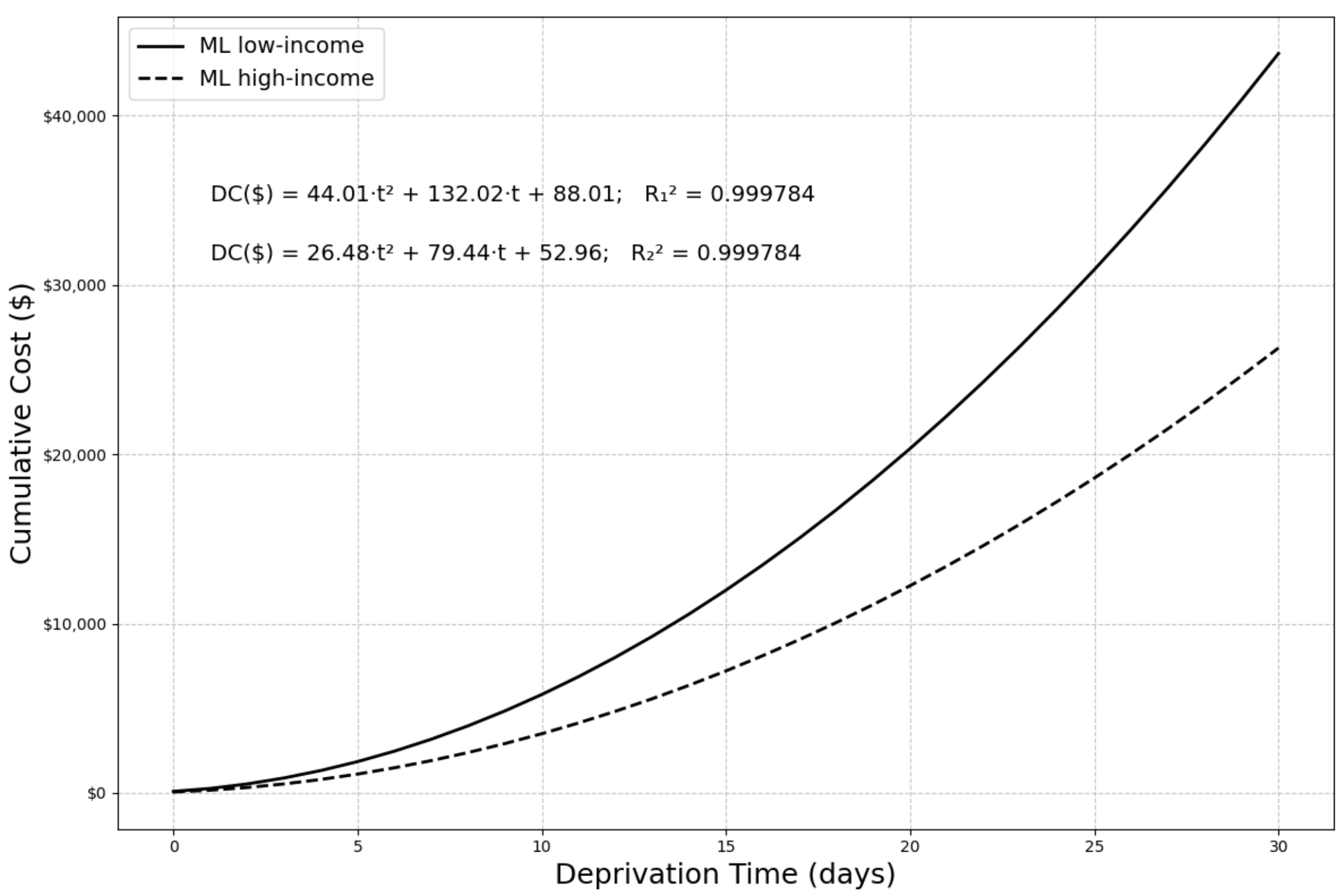

**Figure 7. DCF of MNL1 low-income and MNL1 high-income models.**

## Discussion and Concluding Remarks

This study delivers the first quantitative, time-explicit deprivation cost functions for electrical power outages, addressing a fundamental methodological gap in risk and resilience science. While transportation economics has established methodologies for measuring service disruption impacts[61,62], comparable empirical quantification methods remain underdeveloped for electrical infrastructure[63,64], especially during extreme weather events and other hazardous conditions. This absence of quantified social user costs creates significant challenges in the economic analysis of resilience investments. Grounded in random utility theory and employing discrete choice models with stated-preference data from Harris County, Texas, the study successfully developed DCFs that capture the welfare losses experienced by individuals during power outages. The methodology presents a fully integrated pipeline that converts individual outage-choice data into dollar-denominated, time-explicit deprivation cost functions through four key methodological advances. The discrete-choice framework (i) pivots monetary attributes on each household's actual utility bill to raise consequentiality and external validity, (ii) embeds Box–Cox and exponential time transformations to capture the theoretically convex escalation of suffering, and (iii) estimates different deprivation cost in different income and both systematic and random taste heterogeneity so that household composition, and other socio-demographics are endogenously reflected in the cost curves rather than appended ex-post. The result is a set of empirically grounded coefficients that can be used to calculate a monetary measure of the cost of additional time without electricity—directly compatible with engineering-economic analysis, computable for any scenario horizon, and portable to simulation, optimization, or input-output models of cascading disruption.

Methodologically, the study presents a fully integrated pipeline—from survey design to welfare valuation—that converts individual outage-choice data into dollar-denominated, time-explicit deprivation cost functions. Four elements mark the advance. First, the choice experiment pivots the price attribute on each respondent's actual monthly bill, a consequential-design tactic that sharply reduces hypothetical bias and—unlike prior deprivation work—anchors willingness-to-pay on real household budgets. Second, the econometric architecture nests multiple layers of heterogeneity and non-linearity in a single estimation step: mixed logit models capture unobserved taste variation via panel effects, while Box-Cox and exponential transformations allow the deprivation-time coefficient itself to curve endogenously, yielding statistically defensible convex DCFs without imposing a priori functional shapes. Third, we estimate models for different income groups and analyze systematic interactions (children) directly in the utility specification, thus producing subgroup-specific cost curves that can be fed straight into equity-oriented optimization or scenario tools. Finally, the study demonstrates an explicit welfare-integration procedure: marginal rates of substitution are analytically integrated (or discretely summed) over time to generate complete cost curves that can plug into benefit–cost, input–output, or network-recovery models without further calibration. Together, these innovations deliver a transferable methodological blueprint for quantifying user costs of any infrastructure service interruption—not just power—but also water, transit, or communications, thereby extending deprivation analysis from humanitarian logistics into mainstream resilience economics.

By supplying a dollar-denominated cost term, the project recasts how utilities, regulators, and emergency-management agencies evaluate resilience options. Grid-hardening initiatives, rapid-restoration protocols, and distributed-energy incentives can now be appraised not only on avoided repair expenses but also on the steep, time-sensitive welfare losses they prevent—often an order of magnitude larger for low-income or high child-presence households. Incorporating DCFs into standard benefit–cost or multi-criteria decision tools thus shifts project rankings toward interventions that deliver the greatest aggregate social welfare and the greatest equity dividends. Moreover, because our functions are convex, they mathematically reward speed: every hour shaved from long outages yields a disproportionately higher social benefit than the hour before, providing a clear economic rationale for "first-restore-the-most-vulnerable" dispatch rules. In sum, the study equips decision-makers with an operational metric that (1) quantifies hidden social harm, (2) justifies proactive investment, and (3) targets relief resources where marginal benefits are highest, fundamentally transforming resilience planning from a damage-centric to a people-centric enterprise.

While this study provides a significant contribution, certain limitations should be acknowledged. The findings are based on stated-preference data, which, despite careful survey design, may be subject to hypothetical bias. The geographical focus on Harris County, Texas, means that the specific DCF values may not be directly transferable to other regions or countries with different socioeconomic, demographic, or infrastructural characteristics without further validation. The scope of power outage scenarios, while systematically designed, cannot encompass all possible real-world variations. Future research should aim to extend this work in several directions. Comparative studies across different geographical contexts, disaster types, and for various critical infrastructure interdependencies would enhance the generalizability and applicability of DCFs. Further investigation into the psychological and behavioral factors influencing individuals' perception of deprivation could enrich the utility function specifications. Longitudinal studies, where feasible, could provide insights into how deprivation costs and coping mechanisms evolve over the course of prolonged or repeated disruptions. Finally, a critical next step involves the practical integration of these empirically derived DCFs into operational decision-support tools and policy frameworks for infrastructure management, emergency response, and the development of community resilience strategies, ensuring that the societal cost of deprivation is a central consideration in safeguarding essential services.

## Data Availability

The data that support the findings of this study are available from Qualtrics, but restrictions apply to the availability of these data, which were used under license for the current study. The data can be accessed upon request. Other data we use in this study are all publicly available.

## Code Availability

The code that supports the findings of this study is available from the corresponding author upon request.

## Acknowledgement

This material is based in part upon work supported by the National Science Foundation under Grant CMMI-1846069 (CAREER). The authors also would also like to acknowledge the data support from Qualtrics. Any opinions, findings, conclusions, or recommendations expressed in this material are those of the authors and do not necessarily reflect the views of the National Science Foundation and Qualtrics.

## Author Contributions

All authors contributed to the final manuscript and approved its submission. X.L. is the first author and was responsible for main survey design, data collection, conducting the primary analysis, interpreting key findings, and drafting the manuscript. M.A. was responsible for the main survey design, part of function design, manuscript revision. R.W. was responsible for part of survey design, manuscript revision. B.L. was responsible for part of figure design, and introduction. A.V. was responsible for manuscript revision. A.M is the faculty advisor, guiding the project’s conceptual development, overseeing manuscript revisions, and providing critical feedback on the research.

## Competing Interests

The authors declare no competing interests.